\documentclass[conference]{IEEEtran}
\IEEEoverridecommandlockouts
\usepackage{cite}
\usepackage{tabularx}
\usepackage{amsmath,amssymb,amsfonts}
\usepackage{algorithmic}
\usepackage{graphicx}
\usepackage{textcomp}
\usepackage{xcolor}
\usepackage{pifont}
\usepackage{multirow}
\usepackage{subcaption}
\usepackage{indentfirst}

\usepackage{capt-of}
\usepackage{varwidth}

\def\BibTeX{{\rm B\kern-.05em{\sc i\kern-.025em b}\kern-.08em
    T\kern-.1667em\lower.7ex\hbox{E}\kern-.125emX}}
\begin{document}

\title{TIMBER: On supporting data pipelines in Mobile Cloud Environments}

\author{
\IEEEauthorblockN{Dimitrios Tomaras, Michail Tsenos, Vana Kalogeraki}
\IEEEauthorblockA{
Athens University of Economics and Business,\\ 
Athens, Greece\\
\{tomaras, tsemike, vana\}@aueb.gr}
\and
\IEEEauthorblockN{Dimitrios Gunopulos}
\IEEEauthorblockA{National and Kapodistrian University of Athens\\
Athens, Greece\\
dg@di.uoa.gr}
}

\maketitle

\begin{abstract}
The radical advances in mobile computing, the IoT technological evolution 
along with cyberphysical components (e.g., sensors, actuators, control centers)
have led to the development of smart city applications that generate raw or pre-processed
data, enabling workflows involving the city to better sense the urban environment and 
support citizens’ everyday lives. 
Recently, a new era of Mobile Edge Cloud (MEC) infrastructures has emerged 
to support smart city applications that aim to address the challenges raised due to the 
spatio-temporal dynamics of the urban crowd as well as bring scalability and 
on-demand computing capacity to urban system applications for timely
response. In these, resource capabilities are distributed at
the edge of the network and in close proximity to end-users,
making it possible to perform computation and data processing
at the network edge.
However, there are important challenges related to real-time execution,
not only due to the highly dynamic and transient crowd, the bursty and
highly unpredictable amount of requests but also due to the resource 
constraints imposed by the Mobile Edge Cloud environment.
In this paper, we present TIMBER, our framework for efficiently supporting mobile daTa processing pIpelines in MoBile cloud EnviRonments that effectively addresses the aforementioned challenges. Our detailed experimental results illustrate 
that our approach can reduce the operating costs by 66.245\% on average and achieve up to 96.4\% similar throughput performance for agnostic workloads.
\end{abstract}


\section{Introduction}

In recent years, smart city applications have gained increasing interest as a mean of enhancing the quality of life for the citizens. A wide range of smart city applications has been developed that address various aspects of the urban environment, such as smart transportation and trip planning systems \cite{tomaras2018modeling}, crowd based traffic information applications \cite{crowdalert,waze}, emergency response systems\cite{kinane2014intelligent}, etc. The plethora of smart city applications has triggered a new era in the research community. For instance, it has been shown shown that bike sharing systems act as sensors of human mobility across the city\cite{tomaras2018modeling}, as they provide an alternative, eco-friendly and entertaining way to commute to destinations across the city. Furthermore, the proliferation of location-based crowdsourcing systems, such as Facebook Places, Glympse, Foursquare's Swarm, Google Share Location and Runkeeper\footnote{https://www.runkeeper.com} in recent years, has also contributed to a new collaboration and sharing era. In these systems, users contribute their locations in real-time, possibly enriched with rich multimedia content, such as images and videos, in order to track family members and friends, get rewards, or receive recommendations about places of interest, and people are motivated to participate in these systems. 
For the vast majority of the individuals there are important benefits participating in these systems, i.e., citizens earning free parking time in smart cities\footnote{http://www.vavel-project.eu/}, friends trying to find each other in busy places such as shopping centers or parks or parents tracking their children locations for safety purposes.
All these are examples of complex, heterogeneous, processing pipelines 
that require the cooperation of cyber ({\it i.e.,} sensors) and software components in order to enable real-time response and support the significant processing efforts required by the aforementioned applications.

Recently, Mobile Edge Cloud (MEC) infrastructures have emerged
to support smart city applications to address the challenges raised due to the spatio-temporal dynamics of the urban crowd as well as bring scalability and on-demand computing capacity to urban system applications for timely response. In MECs, resource capabilities are distributed at the edge of the network and in close proximity to end-users,
making it possible to perform computation and data processing at the network edge.
In these settings, Serverless computing, and in particular Function as a Service (FaaS), has shown great potential as an increasingly popular programming model, fueled by the recent demand to host services on provisioned cluster infrastructures and the paradigm shift towards supporting interconnected processing pipelines in today's environments.
The serverless computing model offers an intuitive, event-based interface for developing cloud-based applications, that makes the writing and deployment of scalable microservices both easier and cost effective. 
The serverless computing model has been successfully adopted
in a wide range of application domains, such as for ingesting IoT data for flood warnings\cite{leal2023cloud} and for air quality monitoring and alerting in smart ports \cite{ortiz2022microservice}. 
There are open-source serverless deployments such as OpenFaas, Apache OpenWhisk  and 
KNative (https://knative.dev/), and commercial ones such as 
AWS Lambda (https://aws.amazon.com/lambda/), Google Cloud Functions (https://cloud.google.com/functions) and Azure Functions (https://azure.microsoft.com/en-gb/products/functions/).
Its many advantages include: (a) lower deployment costs where many of the management tasks are taken care by the platform provider and as a result users do not explicitly provision or configure virtual machines (VMs), 
(b) resource elasticity where applications can scale up to tens of thousands of cloud functions on demand, in seconds, with no advance notice, 
and (c) lower operational costs based on a pay-as-you-use policy where users only get charged based on the number of resources consumed by the application functions during execution\cite{elgamal2018costless}.
Recent research has shown, that, serverless architectures can be successfully utilized to support the execution of data intensive mobile applications\cite{jiang2021towards} and mobile data processing pipelines with high demand for data parallelism\cite{el2021microservices,leal2023cloud}. 

So what makes resource provisioning for mobile data processing pipelines an essential step? 
{\it First,} the execution of smart city or crowd-sensing tasks is often time-critical. For instance, in an AMBER alert application\cite{romero2021llama,peri2023orchestrating} we need data processing pipelines that can analyze traffic camera feeds across a city in real-time to identify events of interest ({\it i.e.,} individuals or moving vehicles).
Such pipelines have real-time response demands within which the results should have been received. Real-time execution is a challenging problem in these settings, not only due to the highly dynamic and transient crowd, but also due to the resource constraints imposed by the Mobile Edge Cloud environment. 

Second, executing a serverless function requires the function code (e.g., user code, language runtime libraries) to be brought from persistent storage into main memory (a phenomenon known as {\it cold-start}). Keeping the functions in memory at all times may be prohibitively expensive for the emergency service provider, as function calls can be very sparse or other times highly dynamic. {\it Cold start} refers to the set-up time required by the FaaS provider to get a serverless function’s environment up and running before executing the function. 
The cold start time can be a significant fraction of the function's execution time and rises sharply with an increased but unpredictable number of function requests \cite{silva2020prebaking}. Furthermore, the functions can vary widely with respect to their resource needs and invocation frequencies from multiple triggers, making the prediction of function invocations a rather challenging problem. 
As a result, the high cost of the container startup process makes it extremely challenging for FaaS providers to deliver high elasticity services for mobile data streams in Mobile Edge Cloud environments.

To address the cold-start problem, 
recent research has proposed i) container optimization techniques \cite{silva2020prebaking,oakes2018sock} 
where the goal is primarily to maintain pools of containers ready for execution, 
and ii) prediction methods \cite{shahrad2020serverless,fuerst2021faascache}
that aim to estimate the number of resources needed. 
However, both approaches are inadequate as they cannot easily deal with highly dynamic and bursty workload or they need to have a significant set of \textit{prewarmed} resources ready for execution.
On the other hand, schedulers built in most widely utilized open-source serverless platforms such as OpenWhisk (https://openwhisk.apache.org/) 
employ locality-based criteria, {\it i.e.,} co-locating invocations
of the same function to a randomly-selected worker without 
taking into consideration load conditions; these techniques are shown to be
ineffective, unable to handle highly-skewed workloads. In \cite{cardellini2012service} the authors have pointed out inadequate support of efficient resource allocation mechanisms by the cloud service providers in order to reduce costs and ensure auto-scalability for two types of traffic: 1) bursty and 2) unpredictable. The authors of \cite{hendrickson2016serverless} have proved that Lambda has solved the first problem of efficient auto-scaling against bursty traffic by sharing a pool of containers among multiple instances of different applications. However, such approaches, are still limited because they fail to consider mobile data streams characteristics, such as the type of mobile application using a certain type of data stream, the request rate with which the data will arrive, etc.

In this paper we present our approach for efficiently scheduling the execution of 
mobile daTa processing pIpelines in MoBile cloud EnviRonments.
Our goal is to meet invocation rate and execution time requirements for latency-sensitive mobile applications with resource and monetary cost efficiency.
Our work advances state-of-the-art methods as we address the resource provisioning problem for mobile data stream applications, with zero \textit{a priori} knowledge, which has not been exploited by previous works.
In cases that knowledge from past runs is not available to estimate the amount of resources and make load predictions, we utilize a graph similarity approach to capture the similarities among serverless applications, exploiting the graph edit distance metric\cite{zeng2009comparing}, which has exhibited superior performance compared to alternative techniques (we also illustrate this in our experimental evaluation section) to ensure the scalability of our proposed approach.

In our work we make the following contributions:
\begin{itemize}
    \item We propose TIMBER, our framework that can efficiently support the execution of mobile daTa processing pIpelines in MoBile cloud EnviRonments (MECs).
    
    \item We build TIMBER by employing a neural network prediction model that allows us to predict efficiently the degree of function replication as well as select the appropriate configuration to satisfy real-time deadlines, while minimizing operating costs.
    \item We exploit the idea that similar processing pipelines share certain properties ({\it i.e.,} execution time), 
    and thus we use the appropriate prediction model for pipelines that exhibit similar function call graphs
    via a Graph Edit Distance (GED) metric and derive appropriate configurations that satisfy 
    user throughput and pipeline completion time constraints, even for processing pipelines \textit{with zero a priori knowledge}.
    \item We have implemented our prototype on top of Apache Mesos and Mesosphere Marathon and evaluate TIMBER across different processing pipelines based on real datasets.
    \item Our detailed experimental results illustrate the working and benefits of our approach, which can reduce the operating costs by 66.245\% on average and achieve up to 96.4\% similar throughput performance for agnostic workloads.
\end{itemize}

\section{System Model and Problem Definition}

\subsection{System Model}

{\bf Mobile Data Pipelines.} We utilize the serverless computing model, where the mobile data processing pipeline code is deployed at the granularity of functions which are invoked upon request. The function abstraction provided by the serverless computing model by its design allows to build mobile data pipelines of subsequent functions,
where each function has a specific processing functionality (e.g. run kmeans on top of a sample of GPS data, run an AES encryption algorithm on them, store the encrypted data to an external storage, etc.), thus enhancing the reusability of the same code segments from multiple processing pipelines. In this paper, we consider $k$ heterogeneous serverless functions that are hosted on a serverless environment and comprise the mobile data pipelines. More formally, let $\mathcal{F}:\{f_1,...,f_k\}$ denote this set of $k$ heterogeneous serverless functions hosted in this environment. The number of active instances (also known as replicas, we use the terms function replicas and container replicas interchangeably) can be specified either by the user or can be adapted dynamically according to the request rate. For example, to ensure the timely response of a mobile data processing pipeline, during periods of high
load, the number of active instances can be adapted automatically
to adjust to the increased traffic. On the other side, during
extended periods of inactivity, the number of active instances can
decrease to zero (scale to zero) to keep the total execution cost
low. 
Let $|r_{f_k}^{j}|$ denote the number of replicas for function $f_k$ instantiated in $j$ separate containers, where each $w_j$ container is allocated with $m_{w}$ MBs of memory and $c_{w}$ CPUs. We assume, that, the containers for all replicas of a function $f_k$ are homogeneous, this means that all containers of function $f_k$ have the same CPU and memory allocation. 
The total allocated CPUs and memory for one function is the sum of the CPUs and memory allocated for all replicas of the function. More formally, a container is modelled as follows:
$w_j=\{m_{w},c_{w}\}$
and the configuration of the function $f_k$ (that is, the number of function replicas as well as the memory and cpu allocated for each function replica) is expressed by the following vector
$\overrightarrow{f_k}=\{|r_{f_k}^{j}|,w_j\}$,
where $|r_{f_k}^{j}|$ denotes the number of replicas.
Each function $f_k$ is characterized by its execution time, i.e. the amount of time $\mathcal{T}(f_k)$ it needs to compute. The execution time depends on the algorithmic complexity of the application code and the size of the container $w_j$ (memory $m_w$ and cpu $c_w$), hosting the execution environment of the serverless function\cite{perez2018serverless}.

{\bf Pipeline Completion Time.} Our goal is to optimize the average Pipeline Completion Time (PCT) by predicting the amount of resources required to satisfy a certain rate of incoming mobile data requests.
The Pipeline Completion Time is defined as the total end-to-end time required for a pipeline of serverless functions $f_k$ to complete.
More formally,
\begin{align}
  \mathcal{T}_{pct}(f_k)=\mathcal{T}_{init}(f_k)+\sum^{N}\mathcal{T}(f_k) + qu_k  
\end{align}
The pipeline completion time consists of: \textit{i) the initialization overhead}, $\mathcal{T}_{init}(f_k)$, that is, the amount of time required to 
instantiate all the containers for the execution of the pipeline of serverless functions,
\textit{ii) the queuing time} in the platform queues $qu_k$, and \textit{iii) total execution time required} for the pipeline to run on the Mobile Cloud Environment, that is the total sum of execution times $\mathcal{T}(f_k)$ of each function serverless function $f_k$ of the pipeline.
To complete execution by a service level objective(SLO) deadline $d_k$, we require that
\begin{align}
\mathcal{T}_{pct}(f_k) \leq d_k
\end{align}

\subsection{Problem Definition}
Let a cloud computing platform hosting different pipelines of containerized functions.
Our objective is to determine, for each containerized function, the appropriate configuration of resources that will satisfy the rate of incoming requests while minimizing operating costs.
This corresponds to determining the appropriate \textit{number} (number of replicas $|r_{f_k}^{j}|$) as well as the \textit{configuration} of containers (memory and cpu size) to spawn 
for the specific function, such that it will minimize the pipeline completion time $\mathcal{T}_{pct}(f_k)$, and meet its SLO time deadline, $d_k$. 
The PCT is a linear function of the initialization time, the total execution time for the pipeline to run and the waiting time in the platform queues. The function initialization time is dependent on the orchestrator, the code size and number of replicas of the application, while, the queuing time is related to the total number of functions scheduled for execution. Therefore, the parameter that affects mainly the pipeline completion time and can be further optimized is the execution time of each serverless function to run.

Given a set $\mathcal{W}$ of $i$ possible configurations for pipeline consisting of serverless functions $f_k$, where
$\mathcal{W} = \{\overrightarrow{f_k}^{1},...,\overrightarrow{f_k}^{i}\} =\{(|r_{f_k}^{j_{1}}|,m_{w_{1}},c_{w_{1}}),...,(|r_{f_k}^{j_{i}}|,m_{w_{i}},c_{w_{i}})\}$ 
the problem is to select the appropriate configurations for the serverless functions $f_k$ such that the Pipeline Completion Time $\mathcal{T}_{pct}(f_k)$, will satisfy a certain SLO deadline. This is related to estimating the number of function instances $|r_{f_k}^{j}|$ to spawn that can run in parallel i.e. maximize the probability $\mathcal{P}$ that a set of specific configurations and therefore, a specific number of instances 
can fulfil the user constraints and satisfy the SLO constraint. More formally, the problem can be formulated as:
\begin{align}
    max \mathcal{P}(\mathcal{W}=\{\overrightarrow{f_k}^{i}\})\\
    s.t. \mathcal{T}_{pct}(f_k) \leq d_k
\end{align}

{\bf NP-Hard.} We prove that our optimization problem is NP-Hard as it can be reduced from the well-known 0-1 Knapsack problem. Let us consider an instance of our problem. Consider a set of possible configurations $\overrightarrow{f_k}^{i}$ for our pipeline serverless functions $f_k$ that will satisfy our deadline and is associated with a respective monetary cost. We identify this configuration $\overrightarrow{f_k}^{i}$ that maximizes the probability that $\mathcal{T}_{pct}(f_k^{i}) \leq d_k$. Only one of this configurations can be selected. As it can be seen our problem can be reduced to the well-known 0-1 Knapsack Problem. Given an instance of the 0-1 Knapsack problem, we select the appropriate items that maximize the total sum of values and the sum of weights is smaller than or equal to a given budget. If we denote the probability of satisfying the SLO constraint as value and the cost of the configuration as weight, and given that only one configuration is selected, then the 0-1 Knapsack problem is reduced to our problem. Given that 0-1 Knapsack problem is NP-Hard, then our problem is also NP-Hard.

\section{Our Methodology}

In this section we present our approach that uses neural networks for predicting the most appropriate resource configuration for the mobile data processing pipelines in order to satisfy certain SLO deadlines and ensure their timely responses. Then we discuss our methodology for resource provisioning in cases that no a priori information exists for these pipelines.

\subsection{Resource estimation}

To estimate the most appropriate resource configuration, one simple
way is to use grid search, i.e., enumerate all possible combinations
of memory, CPU and number of instances, which would provide the optimal solution. However, the cost of
enumeration is incredibly high since there is a very large number of combinations that need to be explored, and thus this approach is rather infeasible. In the bibliography, there exist approaches like multi-armed bandits\cite{cheng2023bandit} or Bayesian Optimization approaches\cite{peng2019generic}, which are still limited for the setting we consider. Multi-armed bandit approaches require the setup of a set of functions for reward and regret, where, through exploration and exploitation, they will likely derive a sub-optimal configuration. On the same path, Bayesian Optimization approaches, through sampling and minimizing the number of trials to identify a near-optimal solution \cite{bilal2023great}, require also a trial-and-error approach in order to find the solution. 

Our intuition comes from a different perspective: Can we exploit pre-existing knowledge from other pipeline runs in order to derive the best configuration for the serverless functions of a specific pipeline? \textit{Transfer learning} refers to a technique for predictive modeling on a different but somehow similar problem that can then be reused partly or on its entirety to accelerate the training or improve the performance of a model on the problem of interest. This characteristic can be quite beneficial if exploited appropriately. 

In TIMBER, we opt for a prediction model approach. More specifically, several works in the literature have utilized neural networks as the appropriate prediction model for different types of inference tasks. Typically, in a neural network setting, an agent learns how to benefit most from making sequential decisions by iteratively propagating information back and forth during the training phase and accumulating knowledge from previous experience. This fundamental characteristic is inline with our intuition that existing knowledge can serve as the appropriate means for enhancing a prediction model.

To this end, in TIMBER, we build a Sequential Neural network model for predicting the appropriate configuration for each one of the serverless functions comprising the mobile data processing pipeline. A Sequential model\cite{donkers2017sequential} by its design consists of a plain stack of layers where each layer has exactly one input tensor and one output tensor i.e. it allows us to build a model by stacking layers of nodes (neurons) on top of each other. The Sequential model is the simplest neural network base model in Keras (https://keras.io/). 
Each argument of the Sequential constructor is a layer of neurons; in this case Dense layers. In dense layers (or densely-connected or fully-connected) all the neurons receive an input from all the neurons present in the previous layer.

{\bf Input Layer.} 
Each neuron has an activation function which computes the value that is passed on to the neurons in the next layer. In terms of the layers in TIMBER we choose the ReLU function as an activation function in all layers apart from the last one, which has shown faster convergence times as shown in various works in the bibliography\cite{krizhevsky2012imagenet}, as it requires the estimation of a max value in each neuron rather than the estimation of exponential formulas compared to the sigmoid activation function. More formally, the activation function is described as:
$g(h)=h^{+}=max(0,h)$
where $h$ is the input to a neuron. The input layer takes into consideration the container allocation, {\it i.e.,} the CPUs and memory allocated for each of the functions of the pipeline as well as the request rate of the pipeline. We use the request rate of the pipeline to estimate the pipeline completion time (PCT) to satisfy a certain deadline, as requested from the service provider.

{\bf Output Layer.} The goal of the neural network is to predict the appropriate number of function replicas for each function of the mobile data processing pipeline required in order to satisfy the specific request rate, as defined by the service provider. In TIMBER, the number of predicted necessary instances that will satisfy the user imposed constraints is translated to a set of labels, where each label refers to one of the serverless functions that consist the pipeline.  At the very last step of our prediction algorithm using the neural network, it is required to normalize its output to a probability distribution over predicted output classes. For this purpose, in the last layer of our neural network, we utilize the softmax activation function, based on Luce's choice axiom\cite{luce1977choice}. The softmax activation function implies that we have different probabilities among the different labels. More formally, this implies that
$\sigma(\overrightarrow{\rm q_{n}})_{n}=\frac{e^{q_{n}'}}{\sum_{b=1}^{K}e^{q_{b}'}}$,
where $\sigma(\overrightarrow{\rm q_{n}})_{n}$ represents the prediction probabilities for each one of the possible labels, $K$ is the number of classes in the multi-class classifier and $q_{n}$ are the elements of the input vector to the softmax function, and they can take any real value, positive, zero or negative. For example a neural network could have output a vector such as (-0.62, 8.12, 2.53), which is not a valid probability distribution, hence why the softmax would be necessary.

{\bf Loss Function.}
A prerequisite of the model training process is to monitor how well the model's prediction fits the training data. This is denoted by the loss (or cost) function of a model, where the target is to minimize the loss value by adjusting the weights accordingly. In TIMBER, we use cross-entropy \cite{zhang2018generalized}, 
a widely used loss function when optimizing classification model. In order to speed up the training process, we opt for using the cross-entropy error instead of the sum-of-squares error function, as well as, to improve the generalization of the model\cite{gordon2020uses}. 
Another important aspect in our prediction problem setting is that we aim to solve a multiple class classification problem. For this reason, we opt for categorical cross-entropy rather than binary cross-entropy (since we consider each number of instances as a different label). In our experimental evaluation, we show extensively that other metrics such as the Kullback–Leibler divergence\cite{cao2020deconvolutional} 
(which indicates that the two data distributions in question have identical quantities of information) and the Poisson distribution\cite{magill2018neural} (which is a generalized linear model form of regression analysis used to model count data), are outperformed by the categorical cross-entropy metric.

\subsection{Pipeline Similarity}

Reinforcement learning models have shown to be a good fit\cite{yu2021faasrank} for learning policies for computer systems, because the model agents are capable of learning from real-world workloads and operating conditions without human-designed inaccurate assumptions and interference. More specifically, the model learns how to benefit most from making sequential decisions by iteratively interacting with the environment and accumulating knowledge from previous experience. In addition to that, regardless the training process overhead required, neural networks feature an important characteristic: neural networks representations may exhibit significant similarities and correspondences between representations in networks trained from different initializations and they can be identified reliably\cite{kornblith2019similarity}. 
Similar to earlier works \cite{zacheilas2017dione,xin2022locat}, TIMBER assumes that mobile data processing pipelines with similar codebase will have the same behavior for the same size of input, making it an exploitable characteristic to improve the estimation of resource provisioning configurations based on accumulated knowledge of the model from previous experiences, and provide likely similar timely responses.

{\bf Estimating the required resources for mobile data processing pipelines with zero \textit{a priori} knowledge.} Our intuition is that mobile data processing pipelines for different but somehow similar problems can be partially or entirely reused, as the means to estimate the number of instances for each serverless function consisting a new and probably agnostic  mobile data processing pipeline. The main goal is to avoid retraining the neural network model for each different pipeline and overcome any limitations regarding the size of the trained model.
Recent works in the literature\cite{zacheilas2017dione} have exploited the notion of execution plans of distributed VM-based applications, rather than mobile data processing pipelines consisting of serverless functions. On the other hand, we opt for the finer-graind notion of call graphs\cite{obetz2019static} for estimating the similarity between mobile data processing pipelines. 
Each mobile data processing pipeline comprises a sequence of serverless function calls. Our goal is to exploit that different pipelines may have similar serverless functions call graphs.

To compute the similarity between the call graphs comprising the serverless applications we need a graph similarity measure. There have been multiple graph similarity metrics but with the most popular to be: \textit{i) the graph edit distance ($GED$)} \cite{zeng2009comparing,zacheilas2017dione}, \textit{ii) the maximum common subgraph ($MCS$)}\cite{bunke1998graph} and \textit{iii) the Prefix Preference} \cite{lai2023modelkeeper}.
We decided to opt for the Graph Edit Distance metric as it has been accepted as the most appropriate measure for representing the distance between graphs. More specifically, $GED$ defines the similarity between two graphs by the minimum amount of required distortions to transform one graph into the other. Moreover, $GED$ is error-tolerant and can identify similar graphs even in the presence of noise and errors. \textit{Prefix preference} on the other hand, though it has been adopted in Deep Neural Networks training, like in \cite{lai2023modelkeeper}, does not evaluate the whole mobile data processing pipeline, and mostly, it focuses on the prefixes of the graphs rather than also the suffixes, as the other two metrics. For this purpose, in our experimental evaluation we report the results for the $GED$ and the $MCS$  metrics.
For a new mobile data processing pipeline, for which there is no a priori knowledge, we identify which of the existing pipelines have similar serverless function call graphs. Then, we use the trained neural network to estimate the number of instances of each serverless function consisting the pipeline that will satisfy certain SLO deadlines.

{\bf GED Computation.} Let us assume two call graphs, $CG_1$
and $CG_2$ of the mobile data processing pipelines consisting of serverless functions $f_{k}$ respectively and that $ged_{CG_1,CG_2}$ is the $GED$ distance between call graph $CG_1$ and call graph $CG_2$. The main idea is to match the call graph $CG_1$ with exactly the call graph of $CG_2$, 
compute their $GED$s (i.e., the number of
necessary distortions to make two call graphs identical). The lower the values of $GED$, the more similar two call graphs are. 
The main drawback of computing the $GED$ metric is the fact that its computation has exponential complexity in terms of the graph vertices as the problem of measuring the graph edit distance is NP-hard. For this reason we decided to use a well-known approximation technique\cite{zeng2009comparing} that is able
to effectively and in polynomial time approximate the $GED$ between two call graphs. The main idea is to transform a graph structure to a multiset of a special data-structure, called star structure, and then compute the distance between these multisets instead of the actual graphs. Comparing multisets reduces the search space of the problem as they do not consider the complete structure of the original graph.

{\bf Detecting the most similar pipeline.} 
It is necessary to compute the $GED$ of two call graphs in order to detect whether a mobile data processing pipeline exists for which we have already built a prediction model. 
The idea is to compute the $GED$ between the newly submitted pipeline and all the pipelines that comprise the set of mobile data processing pipelines $HQ$ with which the neural network has been built with. We find the pipeline that leads to the maximum $GED$ value and then examine whether this value
is greater than a pre-defined threshold $\mathcal{T}$ 
($T$ is an administrator-defined parameter to control the similarity among two graphs; this can be tuned dynamically based on the degree of similarity we target). 
If this condition is true we simply return the already built prediction model.
In the case that the $GED$ is lower than T then we proceed with the next most similar serverless application in the $HQ$ set.
By performing this step, we can estimate the number of instances for each serverless function comprising the pipeline to \textit{prewarm} and set them ready for execution in order to meet the developer SLO deadline, even if there is no prior knowledge about the pipeline's resource needs or performance.

\section{Implementation}\label{sec:implement}

\begin{figure}[t!]\centering
\begin{minipage}{0.8\linewidth}\centering
\includegraphics[width=\linewidth]{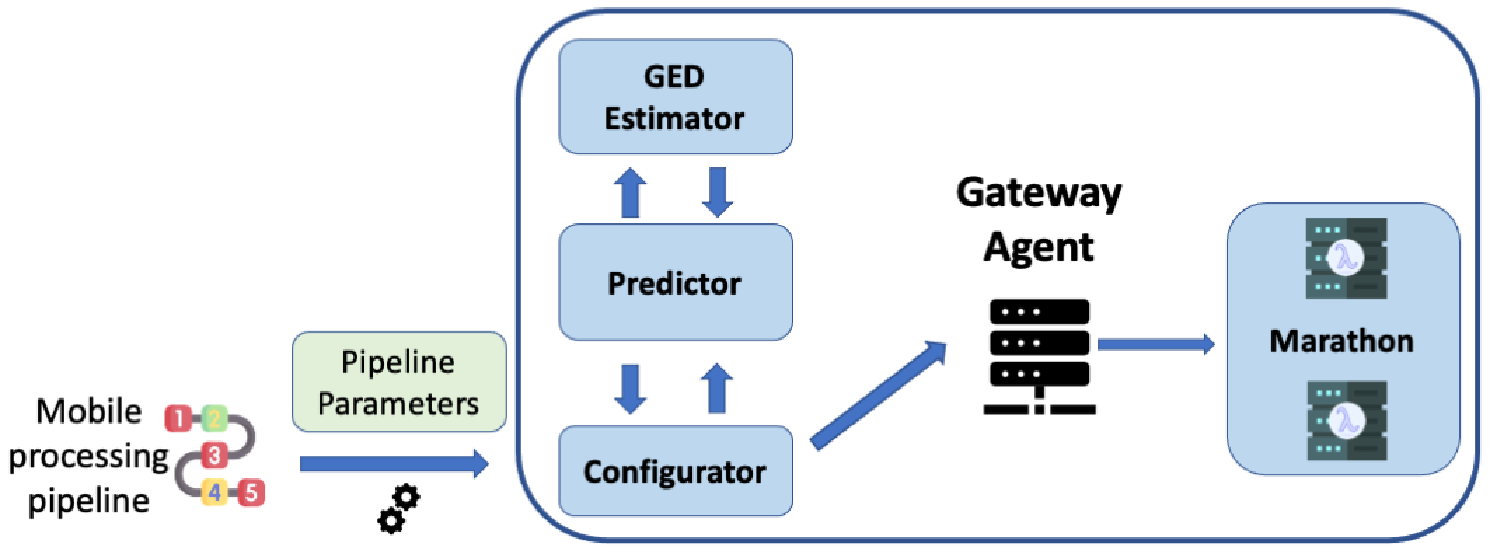}
\caption{TIMBER architecture}\label{fig:overview}
\end{minipage}
\end{figure}

In this section, we discuss the implementation of our TIMBER framework (the TIMBER architecture is shown in Figure \ref{fig:overview}).
TIMBER comprises the following main components: 
(a) {\bf Configurator,} utilized by the developers to define their parameters, i.e.,  
service level objectives for the pipelines and upload the source code for the functions comprising the pipeline. These parameters will be utilized by the Predictor Component to 
specify the number of instances required for each one of the pipeline functions to execute. The functions  will be deployed through our container orchestrator. 
(b) 
{\bf Graph-Edit Distance Estimator,} computes the $GED$ value between two pipelines, that captures the degree of similarity between the developer's submitted pipeline for execution and previously executed pipelines. 
(c) {\bf Predictor:} uses the $GED$ value that was determined by the Graph-Edit Distance Estimator Component in order to estimate the number of instances for each one of the pipeline functions to satisfy the SLO constraint. 
(d) {\bf Gateway Agent.} We developed a Gateway Agent similar to  \cite{tsenos2022amesos} utilized for the deployment of new pipelines as well as to scale up and down deployed serverlesss functions consisting the pipelines. The Agent also acts as a Proxy for function invocations, which are propagated to the deployed function containers. 
(e) {\bf Mesosphere Marathon: } This is our orchestrator (https://mesosphere.github.io/marathon/) that we utilized to start serverless function containers. The deployed containers run in our Apache Mesos cluster. 
Apache Mesos abstracts the compute resources away from machines (physical or virtual), enabling fault-tolerant and elastic distributed systems to easily be built and run effectively. 
Each container listens to port 8080 and Mesos maps that port to a random port of the host Agent. Functions are invoked via sending an HTTP POST request to $http://<agentIp>:<mappedPort>$. All functions are generated with the OpenFaas Function build tools and services that provide the ability to the user to deploy and invoke its functions using the underlying container orchestrator. We adapted OpenFaas to be compatible with Apache Mesos and we used Mesosphere Marathon for easier container deployment on Mesos. We also build our functions using the OpenFaas Watchdog Docker containers, which work as follows: when it receives a request for a pipeline invocation it starts another process within the container that runs the function handler and dispatches the received request to the standard input of the forked process. Then it receives the function result from the function standard output and returns it as the response to the received HTTP request that carried the original function invocation call. Since we do not focus solely on serverless functions but on pipelines consisting of a sequence of serverless function invocations, then the sequence of serverless function call invocations is routed through the gateway agent, since existing production serverless infrastructures do not support \textit{direct} communication between serverless functions explicitly (only through a gateway \cite{gatewaycallfunc}).

\section{Experimental Evaluation}

\subsection{Experimental Setup}
We conducted our experiments in our local cluster comprising 7 nodes (Intel i7-7700 3.6GHz processors), with a total number of 56 CPUs and 112GB of RAM. 
This setup allows us to illustrate the benefits of our approach and balance the trade-off among finding the configuration that satisfies the deadline constraints that is also cost-effective. Similar experimental size setups have been utilized in related works\cite{zacheilas2017dione,jiang2021towards}. 
All the nodes are interconnected with 1Gbps Ethernet. All nodes run on Ubuntu 20.04 LTS. We run Apache Mesos 1.9 as our serverless platform and we use Marathon 1.5 in order to deploy Docker containers on top of Mesos. Marathon is used only to start/stop the containers and no other features are utilized. In this cluster, we deployed our TIMBER system. Similar to related works \cite{jiang2021towards}, we varied the number of serverless function instances from 5 up to 30. We used OpenFaas Python and Java templates in order to create our functions and pipelines and PyCG\cite{salis2021pycg} to extract the call graphs for the pipelines consisting of serverless functions written in Python.

\begin{figure*}[t!]\centering
\begin{minipage}{0.25\linewidth}\centering
\includegraphics[width=\linewidth]{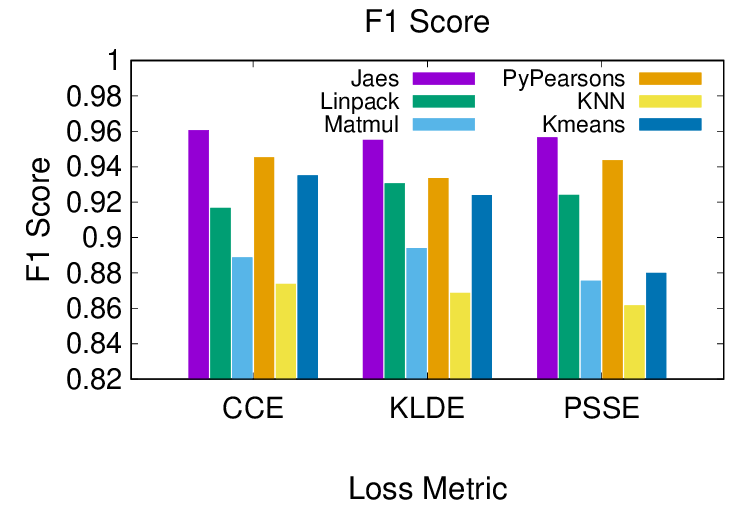}
\caption{F1 Score}\label{fig:f1score}
\end{minipage}\hfill
\begin{minipage}{0.25\linewidth}\centering
\includegraphics[width=\linewidth]{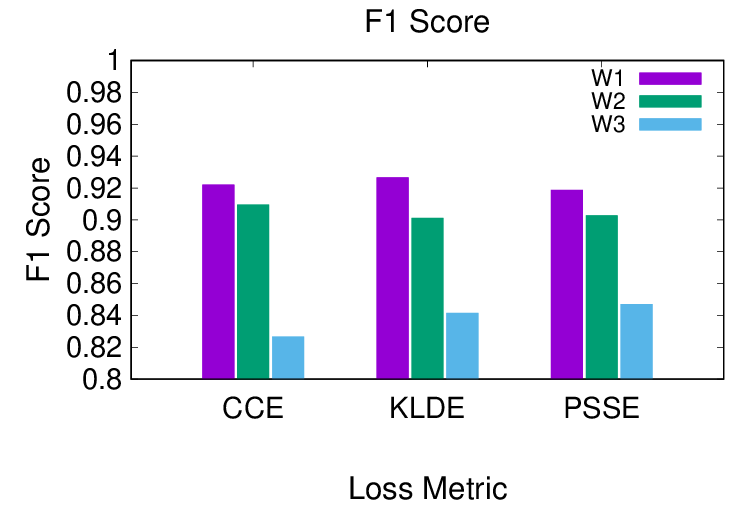}
\caption{Average F1 Score}\label{fig:f1scoreavg}
\end{minipage}\hfill
\begin{minipage}{0.25\linewidth}\centering
\includegraphics[width=\linewidth]{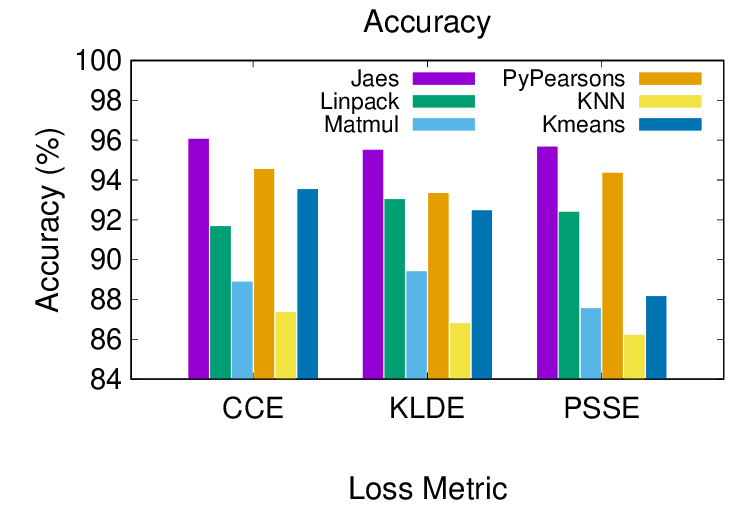}
\caption{Accuracy}\label{fig:accuracy}
\end{minipage}\hfill
\begin{minipage}{0.25\linewidth}\centering
\includegraphics[width=\linewidth]{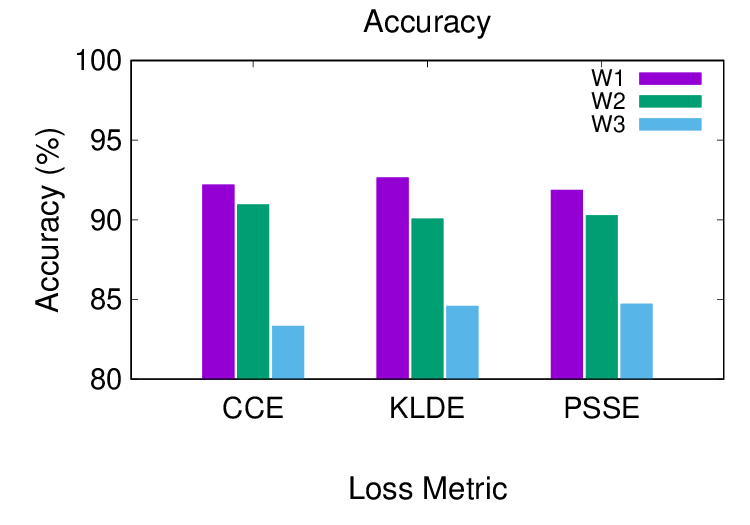}
\caption{Average Accuracy}\label{fig:accuracyavg}
\end{minipage}\hfill
\end{figure*}

\begin{figure*}[t!]\centering
\begin{minipage}{0.25\linewidth}\centering
\includegraphics[width=\linewidth]{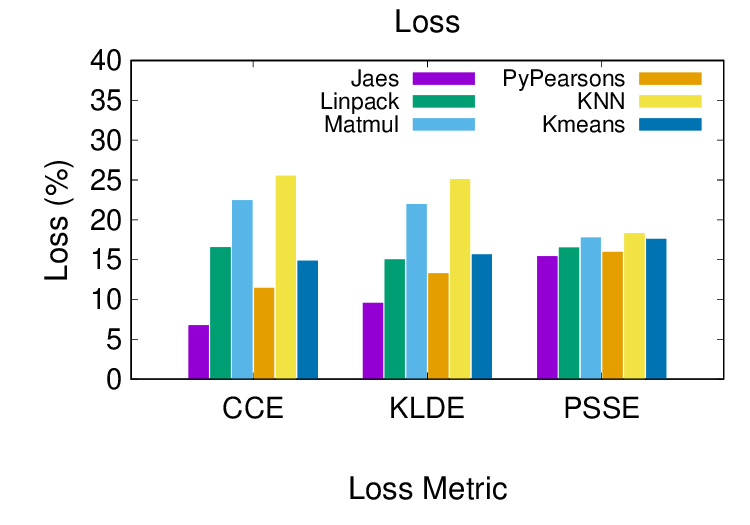}
\caption{Loss}\label{fig:loss}
\end{minipage}\hfill
\begin{minipage}{0.25\linewidth}\centering
\includegraphics[width=\linewidth]{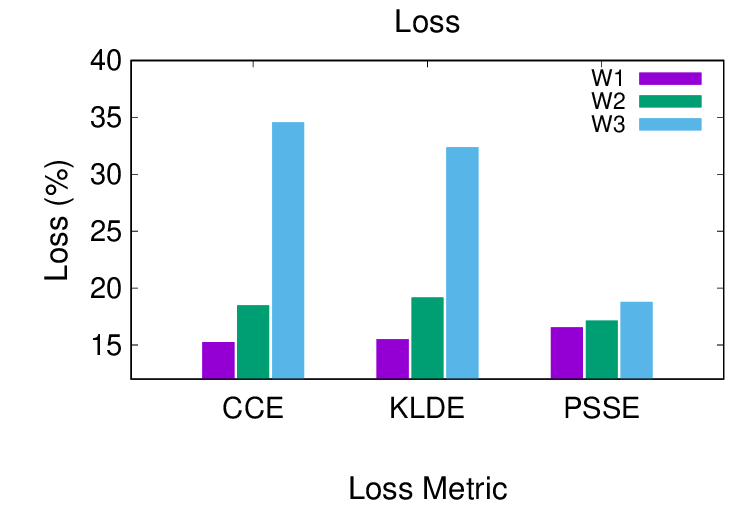}
\caption{Average Loss}\label{fig:lossavg}
\end{minipage}\hfill
\begin{minipage}{0.25\linewidth}\centering
\includegraphics[width=\linewidth]{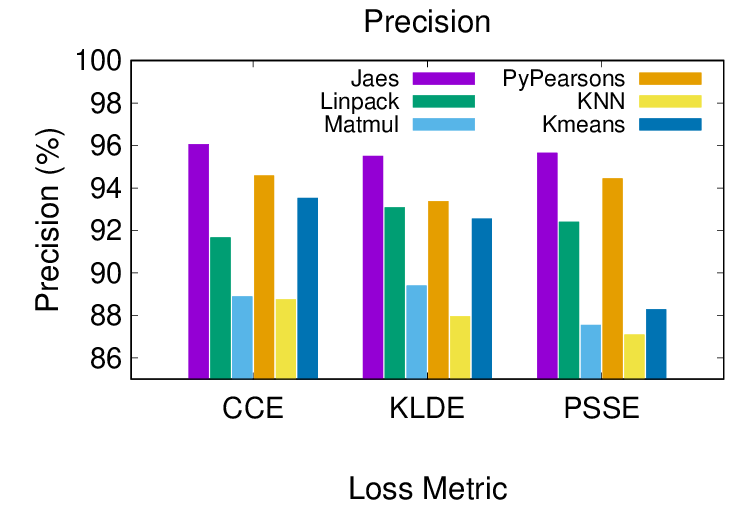}
\caption{Precision}\label{fig:precision}
\end{minipage}\hfill
\begin{minipage}{0.25\linewidth}\centering
\includegraphics[width=\linewidth]{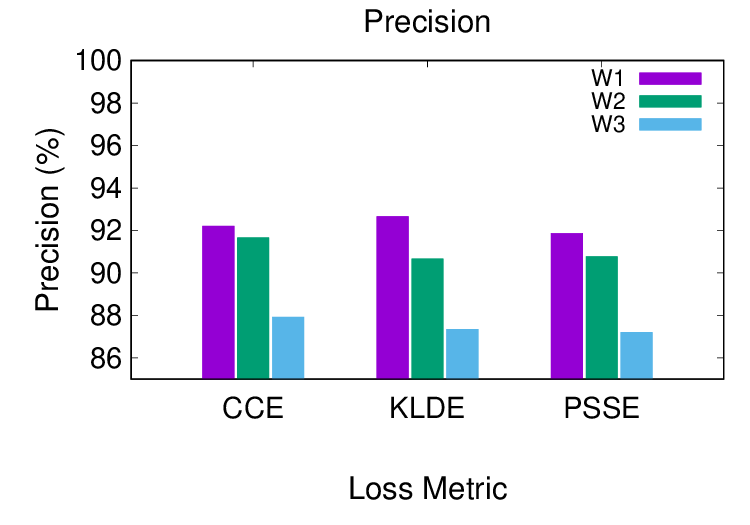}
\caption{Average Precision}\label{fig:precisionavg}
\end{minipage}
\end{figure*}

\begin{figure}[t!]\centering
\begin{minipage}{0.45\linewidth}\centering
\includegraphics[width=\linewidth]{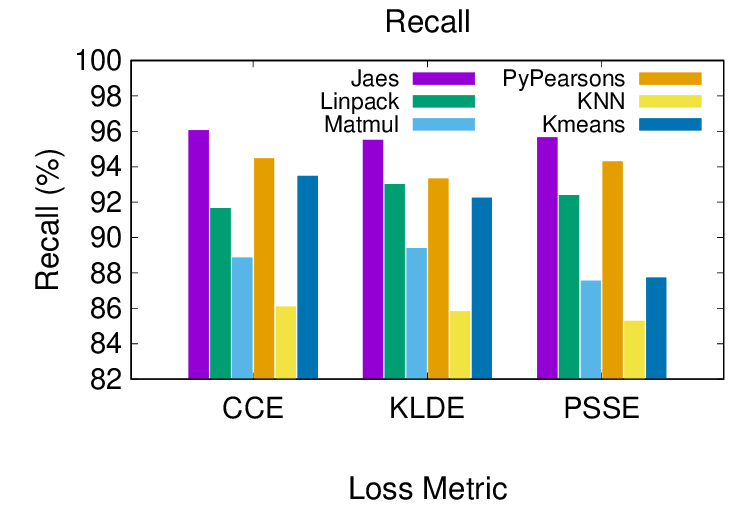}
\caption{Recall}\label{fig:recall}
\end{minipage}
\begin{minipage}{0.45\linewidth}\centering
\includegraphics[width=\linewidth]{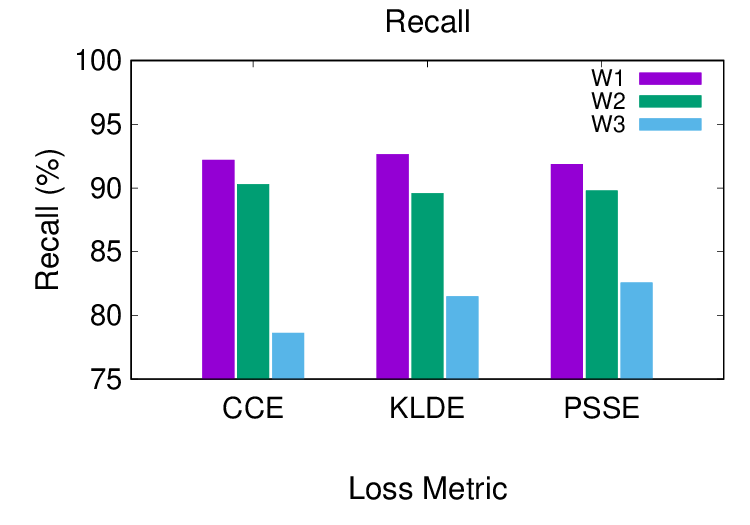}
\caption{Average Recall}\label{fig:recallavg}
\end{minipage}
\end{figure}

\subsection{Functions and Workloads}
{\bf Functions:} In order to evaluate the performance of our approach, we conducted our experiments using real world pipeline scenarios from state-of-the-art performance benchmarks\cite{kim2019functionbench}.
The FunctionBench is composed of micro benchmark and pipeline workload; the micro-benchmark uses simple system calls to measure the performance of resources exclusively, and the pipeline-benchmark represents realistic data-oriented pipeline that generally utilize various resources together. In our experimental evaluation we use six realistic serverless functions that consist mobile processing pipelines chosen from both micro-benchmark libraries\cite{kim2019functionbench}, as well as, developed by us in the context of real-world pipelines as in the works of \cite{zacheilas2017dione,muller2020lambada}. 
Below, we give a brief description for each one of those pipeline functions ({\bf PF}):

\begin{table}[!t]
\footnotesize
    \begin{tabularx}{\columnwidth}{ | c | c | X | }\hline
	    {\bf Workload} & {\bf Pipeline Type } & {\bf Serverless Functions} \\ \hline
	    {\bf W1} & Processing & Jaes, Linpack, Matmul\\ \hline
	    {\bf W2} & Sensor Correlation & PyPearsons, KNN \\ \hline
	    {\bf W3} & Clustering & Kmeans\\ \hline
    \end{tabularx}
\captionof{table}{Different types of pipeline workloads}
\label{tab:serverlapps}
\end{table}

\textit{{\bf PF1} - Jaes: } Jaes benchmark that performs private key-based encryption and decryption. It is a Java implementation of the AES block-cipher algorithm in CTR mode.

\textit{{\bf PF2} - Linpack: } Linpack solves linear equations (Ax = b). We identified that the Linpack serverless function is used by many
state of the art works \cite{manner2021optimizing} to benchmark CPU performance.

\textit{{\bf PF3} - Matmul: } Matmul performs square matrix multiplications. Along with Linpack, these are considered as CPU benchmarks, which are mainly used to measure the CPU-bound performance.

\textit{{\bf PF4} - PyPearsons: } PyPearsons is a Python implementation of Pearsons correlation over a smart-city sensor network and for a given set of geospatial coordinates it returns a list of the most correlated sensors.

\textit{{\bf PF5} - KNN: } KNN is the well known k-nearest neighbor algorithm implemented in Java and for a given set of geospatial coordinates it returns a list of the K nearest sensors from a smart-city sensor network.

\textit{{\bf PF6} - Kmeans: } Kmeans function refers to the Kmeans algorithm as implemented in the scikit-learn python library. Kmeans runs on a small batch of data equal to 2Mb and fits the data to the model.

\begin{figure*}[t!]\centering
\begin{minipage}{0.245\linewidth}\centering
\includegraphics[width=\linewidth]{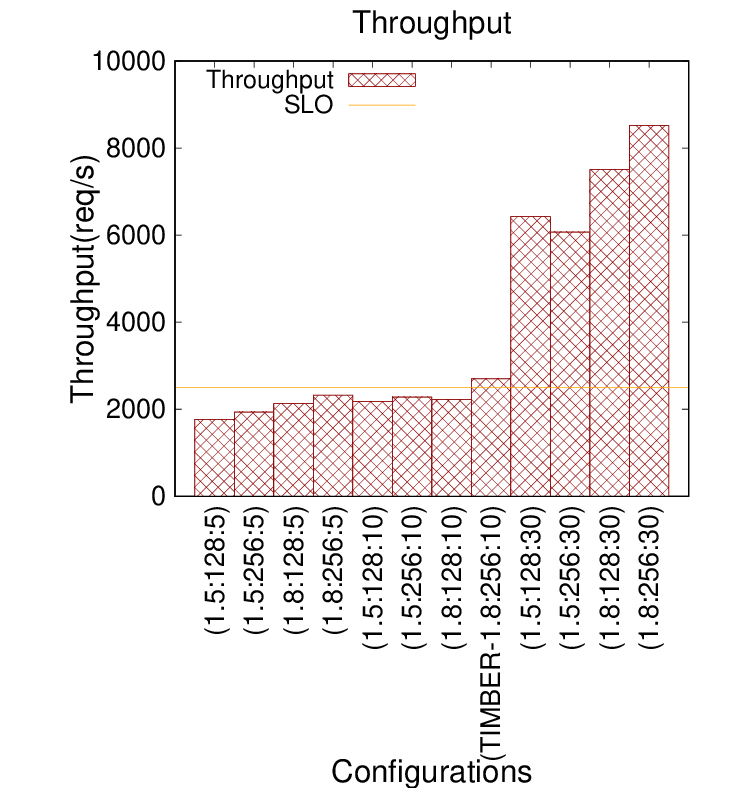}
\caption{Matmul throughput}\label{fig:matthrough}
\end{minipage}\hfill
\begin{minipage}{0.245\linewidth}\centering
\includegraphics[width=\linewidth]{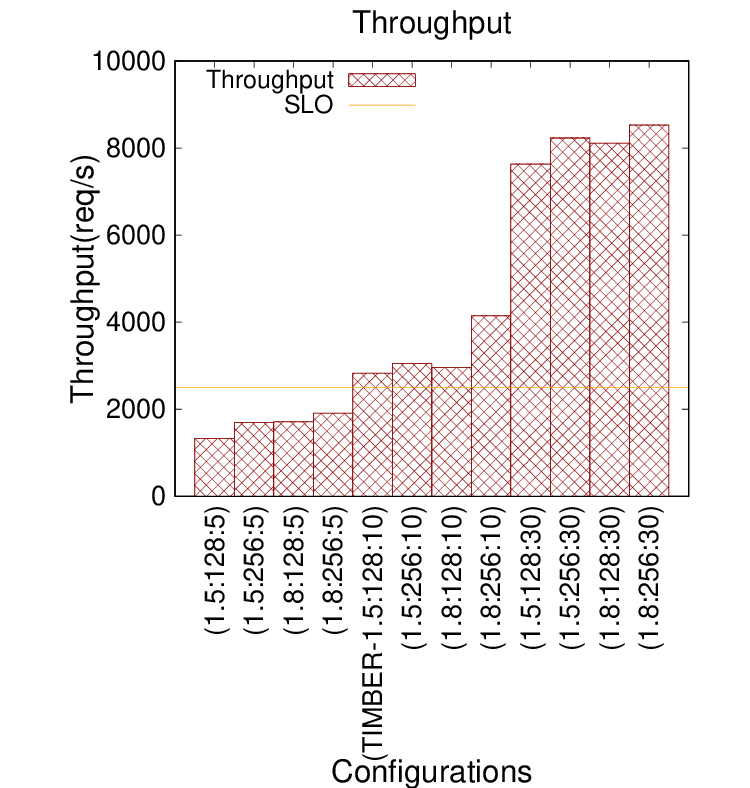}
\caption{Linpack throughput}\label{fig:linthrough}
\end{minipage}\hfill
\begin{minipage}{0.245\linewidth}\centering
\includegraphics[width=\linewidth]{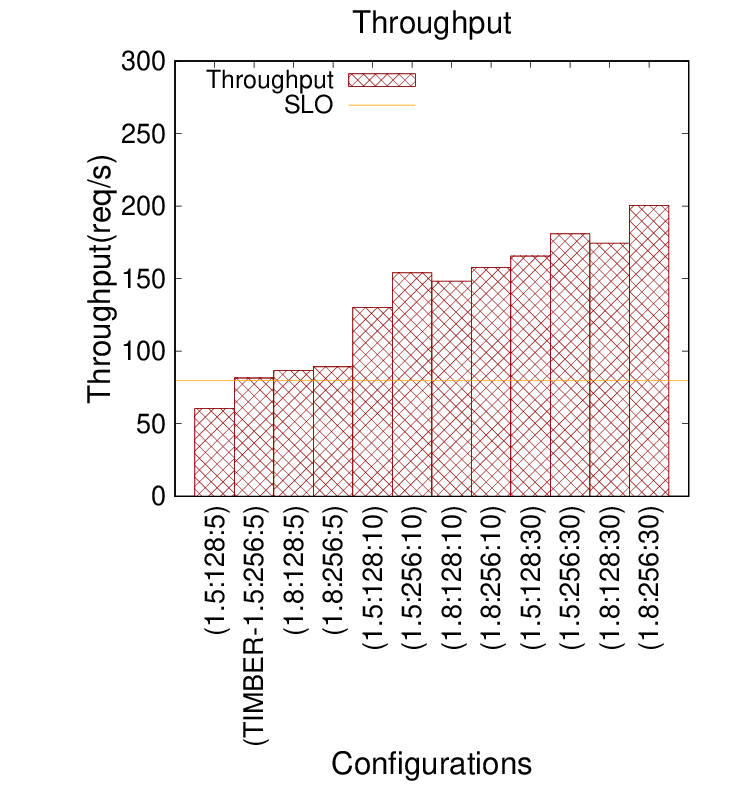}
\caption{PyPearsons throughput}\label{fig:pearsthrough}
\end{minipage}\hfill
\begin{minipage}{0.245\linewidth}\centering
\includegraphics[width=\linewidth]{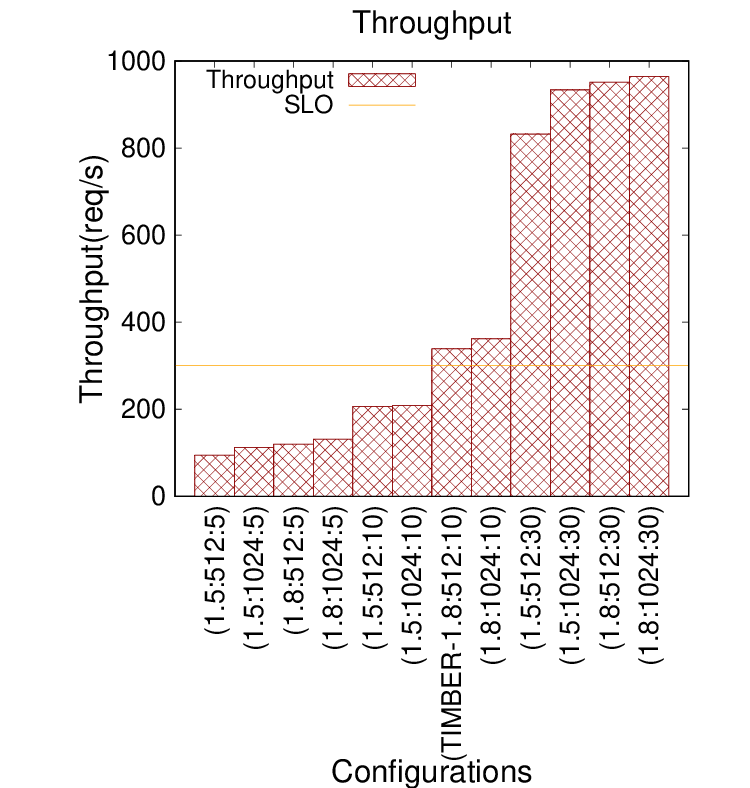}
\caption{Kmeans throughput}\label{fig:kmeansthrough}
\end{minipage}\hfill
\end{figure*}

{\bf Workloads \& Pipelines: } Furthermore, we have grouped these functions in the context of different real-world pipeline types and evaluate their performance, as shown in Table \ref{tab:serverlapps}. This allows us to monitor, identify and understand the behaviour of similar pipelines in the context of a serverless environment. Specifically, these workloads, namely {\bf W1}, {\bf W2} and {\bf W3}, refer to three different types of real-world pipelines: Processing, Sensor Correlation and Clustering respectively. Our initial intuition was to show that similar pipelines with those in the workloads examined will also have similar demands in terms of computational resources as well as the same expected throughput for similar input. Once we have trained the neural model with the pipelines from each one of these different workloads, we can then use it to estimate the number of instances for each other function in similar pipelines and their resource needs even for pipelines for which we have possibly zero knowledge on their performance.

\subsection{Prediction Performance}

\subsubsection{Dataset}
We have evaluated the prediction performance of our neural network utilizing real data from the workloads described in the previous subsection. More specifically, we followed similar guidelines with 
state-of-the-art works \cite{jiang2021towards,zacheilas2017dione}
and utilized four different types of memory and CPU configurations, and also varied the number of replica instances from 5 up to 30. We performed 1000 runs using $Hey$\footnote{https://github.com/rakyll/hey}, for each one of the possible configurations, and aggregated these results in order to construct the training dataset for the neural network deployed in TIMBER.

\subsubsection{Metrics}\hfill

{\bf F1-Score: } The F1-Score metric is computed from the precision and recall of the data. 
The highest possible value of an F-score is 1.0, indicating perfect precision and recall. 

{\bf Accuracy: } The accuracy metric illustrates the frequency with which predicted labels match the true labels. 

{\bf Precision: } The precision metric is the fraction of relevant labels among the retrieved labels, that is the true positive labels over the sum of true positive labels with false positive labels. 

{\bf Recall: } The recall metric illustrates the fraction of relevant labels that were retrieved, that is the true positive labels over the sum of true positive labels and those that were falsely estimated as negative. 

{\bf Loss: } The loss metric refers to the cross-entropy loss (also known as log loss) and captures the performance of a classification model whose output is a probability value between 0 and 1. Cross-entropy loss increases as the predicted probability diverges from the actual label. In TIMBER, we evaluated different types of losses in order to find the appropriate for our problem. These are namely Categorical Cross-Entropy loss\cite{zhang2018generalized}, Kullback–Leibler Divergence Entropy loss\cite{cao2020deconvolutional} and the Poisson Entropy loss\cite{magill2018neural}. We describe our findings regarding those metrics in the following section.

\subsubsection{Findings}
In Figures \ref{fig:f1score}, \ref{fig:accuracy}, \ref{fig:loss}, \ref{fig:precision} and \ref{fig:recall}, we illustrate the behaviour of each different serverless function consisting the pipelines, where in the x-axis we vary the loss metric utilized for the training of the neural model (CCE stands for Categorical cross-entropy loss, KLDE is the Kullback–Leibler Divergence Entropy loss and PSSE is the Poisson Entropy loss). The results show that each different function can achieve very good performance in terms of F1-Score, accuracy, precision and recall, when using the categorical cross entropy as the loss metric required for training. We reason these results due to the fact that KLDE calculates the relative entropy between two probability distributions, whereas cross-entropy can be used to calculate the total entropy between two distributions. That is, CCE can estimate how similar two label distribution are, whereas KLDE how relevant they are. Moreover, we looked into the prediction performance across the different workloads. 
The results illustrated in Figures \ref{fig:f1scoreavg}, \ref{fig:accuracyavg}, \ref{fig:lossavg}, \ref{fig:precisionavg} and \ref{fig:recallavg} validate our initial intuition of choosing the CCE as the appropriate loss metric for our neural network model training. Despite the fact that the loss value of CCE in the case of workload W3 (Clustering) is higher, the overall prediction performance is best when choosing the CCE as the loss metric required for the neural network model training for these three different kind of real-world workloads.

\subsection{Serverless performance}

Please note that due to lack of space, we present the results for the most computationally heavy Python functions (Matmul, Linpack, PyPearsons and Kmeans). 
\subsubsection{Metrics}\hfill

{\bf Throughput: } Throughput is typically defined as the number of requests/second that are successfully served from the serverless system. That is, in TIMBER, our goal is to identify the appropriate container configuration (in terms of cpu, memory allocation) that will achieve a certain level of throughput, as requested from the pipeline developer. 

{\bf Cost: } We evaluated TIMBER's efficiency in terms of monetary units over the duration of one month. Our goal is to illustrate the ability of TIMBER to identify the best configuration in order to meet the developer's defined SLO deadline, while keeping the cost low and flourish the benefits of using a \textit{pay-as-you-use} model.

\subsubsection{Findings}
\begin{figure*}[t!]\centering
\begin{minipage}{0.245\linewidth}
\includegraphics[width=\linewidth]{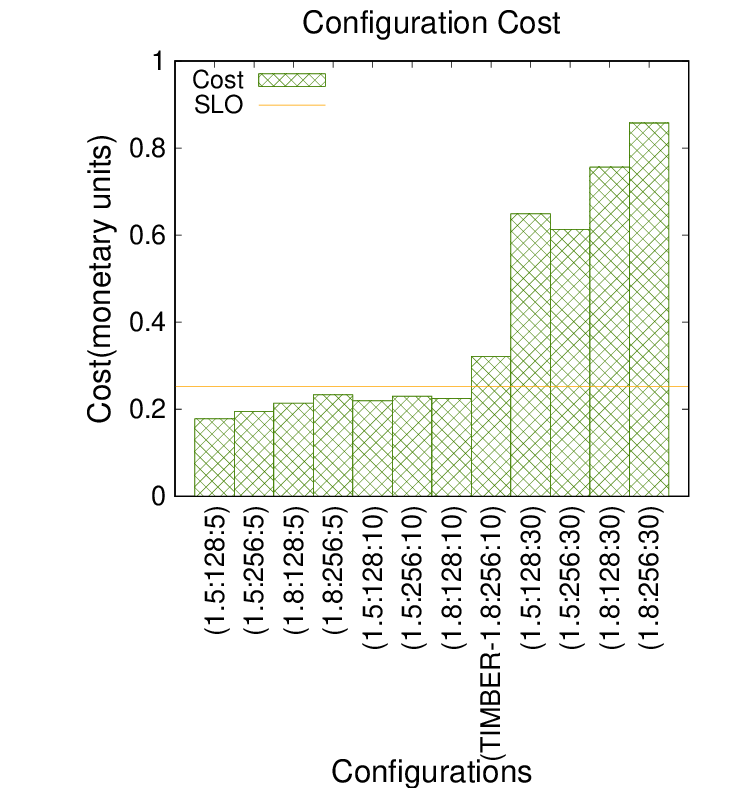}
\caption{Matmul Cost over a month}\label{fig:matmulcost}
\end{minipage}\hfill
\begin{minipage}{0.245\linewidth}
\includegraphics[width=\linewidth]{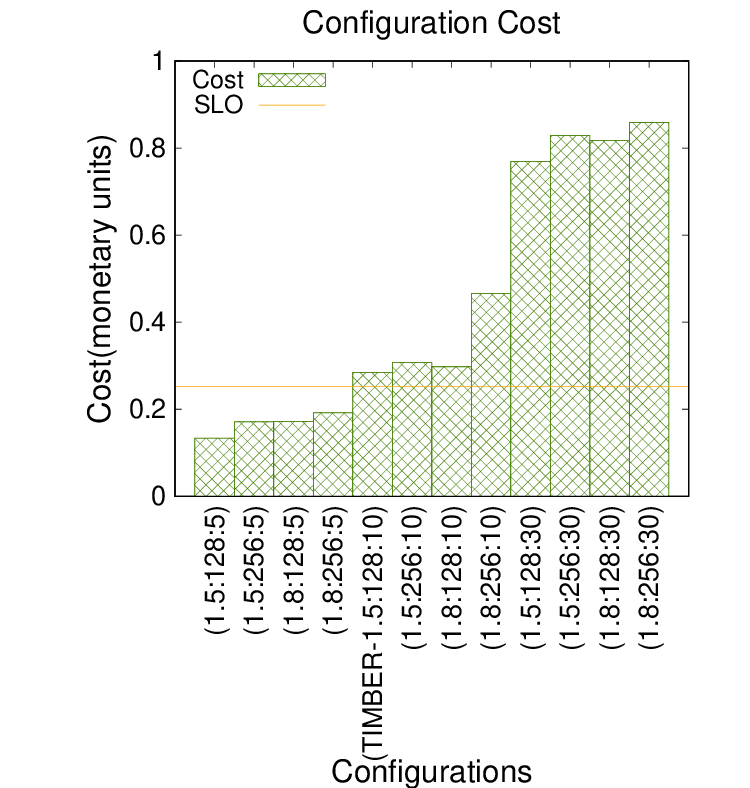}
\caption{Linpack Cost over a month}\label{fig:lincost}
\end{minipage}\hfill
\begin{minipage}{0.245\linewidth}
\includegraphics[width=\linewidth]{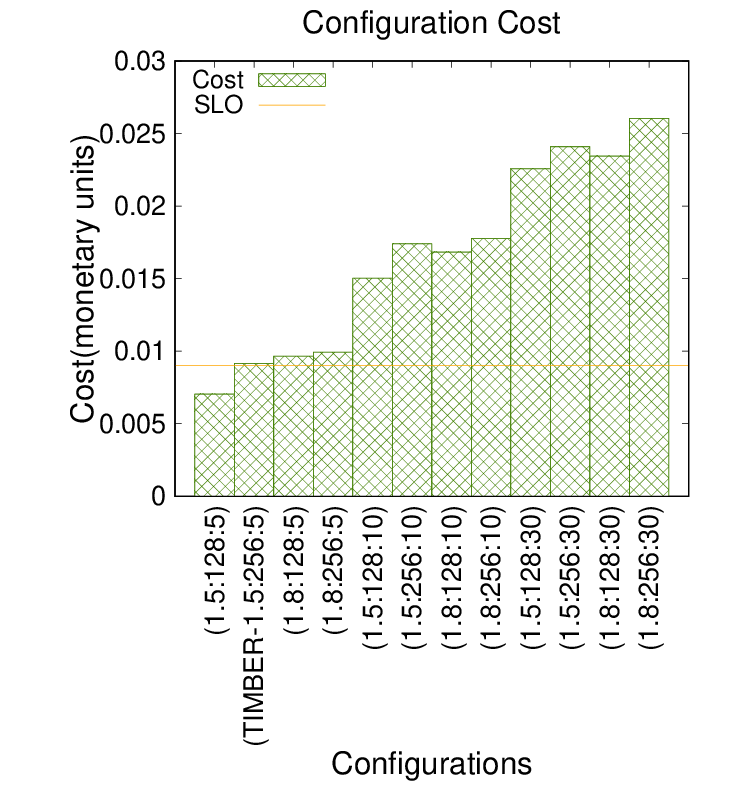}
\caption{PyPearsons Cost over a month}\label{fig:pearscost}
\end{minipage}\hfill
\begin{minipage}{0.245\linewidth}
\includegraphics[width=\linewidth]{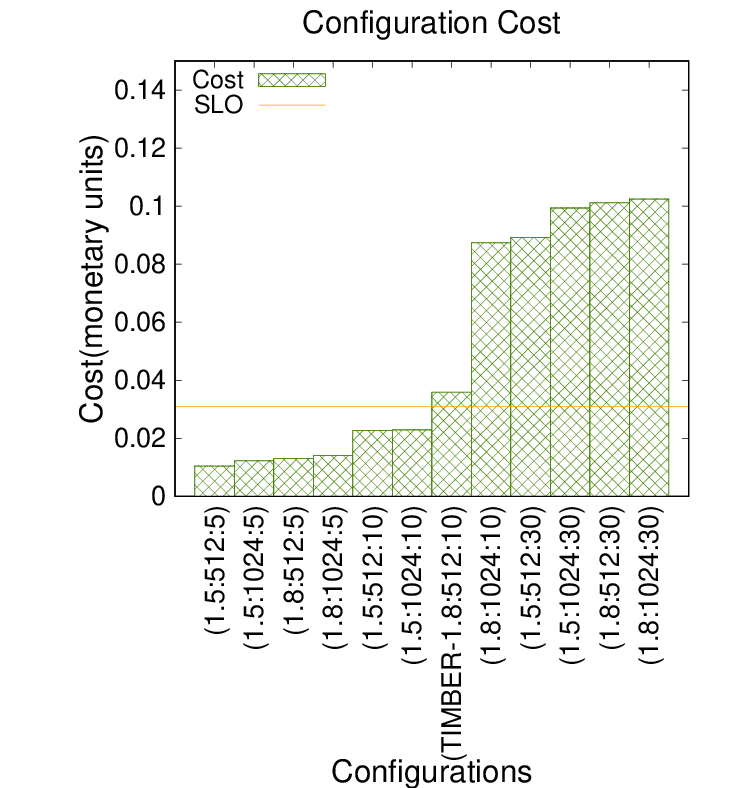}
\caption{Kmeans Cost over a month}\label{fig:kmeanscost}
\end{minipage}\hfill
\end{figure*}
In Figures \ref{fig:matthrough}, \ref{fig:linthrough}, \ref{fig:pearsthrough} and \ref{fig:kmeansthrough}, we draw the throughput achieved by each one of the configurations for a given SLO (equal to 2500 requests / second for matmul and linpack, 80 requests / second for PyPearsons (which is computationally intensive) and 300 requests / second for Kmeans). In the x-axis, we draw all the examined configurations, in y-axis we draw the number of throughput achieved by each one of the examined configuration, and we also annotate the one predicted by TIMBER. We may observe that in every function of each of the three workloads examined (W1, W2 and W3), TIMBER succeeds in estimating the best configuration in order to meet the developer's service level objective. Despite the fact that, there also exist configurations that also succeed in meeting the developer imposed deadline, this may lead to additional provisioning costs, as we discuss next.

In Figures \ref{fig:matmulcost}, \ref{fig:lincost}, \ref{fig:pearscost} and \ref{fig:kmeanscost}, we draw the corresponding costs, using the  pricing scheme from (https://cloud.ibm.com/functions/learn/pricing), for selecting a specific configuration for a pipeline. In the x-axis, we draw all the examined configurations, in the y-axis we draw the cost in monetary units for each one of the examined configurations and we also annotate the cost of the predicted configuration by TIMBER. We may observe that in every function of the three pipelines examined, TIMBER succeeds in predicting and selecting the configuration that not only meets the developer's requirements but is also beneficial in terms of cost (compared to configurations that also succeed in meeting the deadline but require greater number of replicas, thus invoking additional provisioning costs). Compared to choosing naively the greatest number of replicas (i.e. 30) and the largest configuration available, TIMBER can save up to 68.32\% for the Matmul serverless app in terms of operations costs \textit{without overprovisioning} (66.84\% for linpack, 64.86\% for PyPearsons and 64.96\% for Kmeans).

\begin{table}[!t]
\footnotesize
    \begin{tabularx}{\columnwidth}{ | X | c | c | c | }\hline
	    {\bf App} & {\bf GED } & {\bf MCS }&{\bf ps (\%)} \\ \hline
	    {\bf L1} & {\bf GED (PF3,L1)} = 5 & {\bf MCS (PF3,L1)} = 1.5 & 77.81\%\\ \hline
	    {\bf L1} & {\bf GED (PF2,L1)} = 2 & {\bf MCS (PF2,L1)} = 2.0 & {\bf 94.1}\%\\ \hline
	    {\bf L2} & {\bf GED (PF4,L2)} = 2 & {\bf MCS (PF4,L2)} = 1.25 & {\bf 96.4}\%\\ \hline
	    {\bf L3} & {\bf GED (PF4,L3)} = 6 & {\bf MCS (PF4,L3)} = 1.0 & 90\%\\ \hline
	    {\bf L4} & {\bf GED (PF6,L4)} = 2 & {\bf MCS (PF6,L4)} = 1.5 & {\bf 91.9}\%\\ \hline
	    {\bf L5} & {\bf GED (PF6,L5)} = 15 & {\bf MCS (PF6,L5)} = 1.0& 68.1\%\\ \hline
    \end{tabularx}
\captionof{table}{Performance vs GED vs MCS}
\label{tab:gedapps}
\end{table}

{\bf GED vs Estimated Performance: } 
In the last set of experiments we evaluated TIMBER's performance with respect to the estimated configuration for pipelines with zero existing knowledge. We developed 5 different agnostic pipelines ({\bf L1} corresponds to a combination of Matmul and Linpack, {\bf L2} corresponds to a slightly modified version of PyPearsons, {\bf L3} is a moderately modified version of PyPearsons, {\bf L4} is a slightly modified version of Kmeans and {\bf L5} corresponds to an extensively modified version of Kmeans) and computed their throughput using the configuration estimated by TIMBER based on the pipeline they are most similar to. In Table \ref{tab:gedapps}, we summarize 
for each one of the agnostic pipelines,
the corresponding graph similarities with existing pipelines
and the performance similarity $ps$ \cite{kim2020automated}, which is defined as follows:
$ps=|1-\frac{Thr_{Agnostic}-Thr_{SimilarApp}}{Thr_{SimilarApp}}|*100\%$,
where $Thr_{Agnostic}$ is the throughput achieved by the agnostic function and $Thr_{SimilarApp}$ is the estimated throughput by TIMBER for the existing known pipelines. 
We may safely conclude that as the $GED$ gets closer to 1, the performance similarity increases. Similarly, as the MCS metric increases (that is, the two pipelines share multiple same call graph nodes), the performance similarity also increases, which is inline with our findings using the $GED$ metric. Therefore, TIMBER provides good estimations even for agnostic pipelines.

\section{Related Work}

{\bf Container Optimization: } The authors of the paper\cite{silva2020prebaking} focus on how to optimize the container creation by using shortcuts based on checkpoint-and-restore procedures, without the need of recreating the docker container image from the very first step. 
The authors of \cite{oakes2018sock} improve the container boot process to achieve cold starts in the low hundreds of milliseconds.
The work of \cite{shillaker2020faasm} proposes a new lightweight isolation mechanism which, in order to reduce initialisation times, restores Faaslets from already-initialised snapshots.

{\bf Prediction methods: }
The authors of the paper\cite{shahrad2020serverless} focus on reducing the number of cold starts in a serverless environment. 
They propose an adaptive resource management policy called hybrid histogram for prewarming and keep-alive time windows. They also use ARIMA modeling for applications that have infrequent invocations. This is considered as a state-of-the-art technique. However, their work does not consider and exploit the similarity of serverless applications as we do in our work.
The authors of this work\cite{fuerst2021faascache} propose a keep-alive policy based on a Greedy Dual Size Frequency caching scheme, which relies in the container pool of the OpenWhisk serverless platform. The work actually replaces the default TTL scheme which is the typical case of the industry standards. The authors use a function hit-ratio curve for determining the percentage of warm-starts at different server memory sizes. The work of \cite{fu2019edgewise}, is implemented on top of Apache Storm and incorporates a congestion-aware scheduler and a fixed-size worker pool into an edge friendly Streaming process environment, but this could not be applied to our setting, in which we focus on batches of requests.


\section{Conclusions}

In this paper, we presented TIMBER, a framework for estimating the amount of resources required to support different mobile data processing pipelines in a Mobile Cloud environment. We introduced a neural network prediction model that helps transfer learning between different pipelines. We exploited the graph edit distance metric to identify similarities between different but similarly behaving mobile data processing pipelines and derive the appropriate configuration that satisfies certain SLO constraints in terms of throughput and application completion time, even in cases 
of pipelines \textit{with zero a priori knowledge}. Finally, we presented our prototype on top of Apache Mesos and Marathon and evaluated its practicality and applicability based on real datasets, which illustrate a reduction of the operating costs by up to 66.245\% on average and up to 96.4\% similar throughput performance for agnostic workloads.

\section*{Acknowledgment}
This research has been financed by the European Union through the EU ICT-48 2020 project TAILOR (No. 952215), the Horizon Europe AutoFair project (No. 101070568) \& the Horizon Europe CoDiet project (No. 101084642) \& the Hellenic Foundation for Research and Innovation (HFRI) under the 3rd Call for HFRI PhD Fellowships (Fellowship Number: 6812).

\bibliographystyle{IEEEtran}

\bibliography{biblio}

\end{document}